\documentclass[prc,aps,floatfix,twocolumn,groupedaddress,nofootinbib,showpacs,preprintnumbers,superscriptaddress,
amsmath,amssymb,amsfonts,widetable]{revtex4}
\usepackage{array}
\usepackage{hyperref}
\usepackage{xcolor}
\usepackage{graphicx}
\usepackage{bm}
\usepackage{soul}
\usepackage{array}
\usepackage{lipsum}
\usepackage{relsize}
\usepackage{mathrsfs}
\usepackage{amssymb}
\usepackage{amsmath}
\usepackage{graphicx}

\newcommand{\msun}{M_\odot \,}

\newcommand{\Rskin}[1]{R_{\rm skin}^{#1} \,}
\newcommand{\Fskin}[1]{F_{\rm Wskin}^{#1} \,}
\newcommand{\lead}{$^{208}$Pb\,}
\newcommand{\calfe}{$^{48}$Ca\,}
\newcommand{\ca}{$^{40}$Ca \,}
\newcommand{\Ksym}{K_{\rm{sym}}\,}


\begin{document}
\title{Density Dependence of the
       Symmetry Energy\\
       in the Post PREX-CREX Era}
\author{Brendan T. Reed}\email{breed@lanl.gov}
\affiliation{Theoretical Division, Los Alamos National Laboratory, Los Alamos, New Mexico 87545, USA}
\affiliation{Department of Astronomy, Indiana University, 
                Bloomington, Indiana 47405, USA}
\author{F. J. Fattoyev}\email{ffattoyev01@manhattan.edu}
\affiliation{Department of Physics \& Astronomy, Manhattan College,
                Riverdale, NY 10471, USA}
\author{C. J. Horowitz}\email{horowit@indiana.edu}
\affiliation{Facility for Rare Isotope Beams, Michigan State University, East Lansing, Michigan 48824, USA}
\affiliation{Center for Exploration of Energy and Matter and
                  Department of Physics, Indiana University,
                  Bloomington, IN 47405, USA}
\author{J. Piekarewicz}\email{jpiekarewicz@fsu.edu}
\affiliation{Department of Physics, Florida State University,
               Tallahassee, FL 32306, USA}

\date{\today}
\begin{abstract}
 The recently published CREX results suggest a rather peculiar picture for the density dependence of the symmetry energy. 
 Whereas PREX favors a large neutron skin thickness in $^{208}$Pb, thereby suggesting a stiff equation of state, CREX 
  suggests instead a much softer equation of state. This discrepancy has caused a large spur in the theoretical community 
  since no model has been able to simultaneously reproduce within $1\sigma$ the PREX and CREX results. Motivated by a novel correlation 
  between a CREX observable and a combination of bulk symmetry energy parameters, we calibrate three new covariant 
  energy density functionals that reproduce binding energies and charge radii of spherical nuclei---and also accommodate 
  the constraints imposed by PREX and CREX. Given that these models suggest a stiff equation of state at high densities, 
  predictions for neutron star properties are also discussed. 
\end{abstract}
\maketitle

\section{Introduction}
After almost a decade since first conceived, the Calcium Radius Experiment (CREX) at Thomas Jefferson National Accelerator Facility (JLab) 
was recently completed\,\cite{Adhikari:202kgg}. CREX followed on the footsteps of the successful Lead Radius Experiment 
(PREX)\,\cite{Abrahamyan:2012gp,Horowitz:2012tj,Adhikari:2021phr} that aimed to constrain the equation of state (EOS) of neutron-star matter 
in the vicinity of nuclear saturation density by measuring the neutron skin thickness of $^{208}$Pb. Such a powerful connection between 
atomic nuclei and neutron stars is encoded in the nuclear symmetry energy, which quantifies the energy cost of converting symmetric nuclear 
matter into pure neutron matter. In the vicinity of nuclear saturation density, $\rho_0\!\approx\!0.15\,{\rm fm}^{-3}$, the density dependence
of the symmetry energy is parameterized in terms of a few bulk parameters,
\begin{eqnarray}
S(\rho)=J+Lx+\frac{1}{2}K_{\rm{sym}} x^2 +...\ , \quad \, x\equiv\frac{\rho\!-\!\rho_0}{3\rho_0},
\label{eq:sym_energy}
\end{eqnarray}
where $J$ is the value, $L$ the slope, and $\Ksym$ the curvature of the symmetry energy at saturation density. In particular, the correlation 
between the neutron skin thickness of $^{208}$Pb and the radius of a canonical $1.4\msun$ neutron star indicates that 
$L$ controls both the thickness of the neutron skin\,\cite{Brown:2000,Furnstahl:2001un,RocaMaza:2011pm}  and the radius of low-mass neutron 
stars\,\cite{Horowitz:2000xj,Horowitz:2001ya,Carriere:2002bx,Piekarewicz:2019ahf}.

With tiny parity-violating asymmetries of the order of parts per million\,\cite{Adhikari:2021phr,Adhikari:202kgg}, both CREX and PREX-2 determined 
the neutral weak form factors $F_{\rm wk}$ of ${}^{48}$Ca and ${}^{208}$Pb at a single value of the momentum transfer. These values are as model 
independent as the ones extracted decades ago for the corresponding charge form factors $F_{\rm ch}$ using elastic electron 
scattering\,\cite{DeJager:1987qc}. By then exploiting the correlation between the neutron skin thickness 
$R_{\mathrm{skin}}\!=\!R_n\!-\!R_p$ and the weak skin form factor $F_{\mathrm{Wskin}}\!=\!F_{\rm ch}\!-\!F_{\rm wk}$\,\cite{Thiel:2019tkm} 
predicted by several relativistic and non-relativistic models, a constraint on the neutron skin is obtained. Whereas the theoretical model error is small 
in the case of ${}^{208}$Pb (about 15\% relative to the experimental error), the theoretical and experimental errors are comparable for ${}^{48}$Ca. 
Thus, in comparing against experiment it is better to use the model independent CREX weak skin form factor rather than the neutron skin.
Although part of the larger theoretical error in ${}^{48}$Ca is associated to the slightly larger than optimal experimental momentum transfer, ${}^{48}$Ca 
presents additional theoretical challenges relative to ${}^{208}$Pb. For example, the tensor component of the current---often referred to as the 
spin-orbit contribution---becomes important in ${}^{48}$Ca because in a mean field approximation the $f_{7/2}$ orbital is full while its 
$f_{5/2}$ spin-orbit partner is empty\,\cite{Horowitz:2012we}.
Hence, unlike PREX, the connection between CREX and the density dependence of the symmetry energy is not as robust\,\cite{CREX:2013}. 

While some differences between PREX and CREX were anticipated, it came as a surprise to many when CREX reported a central value for the neutron 
skin thickness of ${}^{48}$Ca ($R_{\rm skin}^{48}$) that was significantly smaller than the corresponding value for ${}^{208}$Pb ($R_{\rm skin}^{208}$).
Although the correlation between $R_{\rm skin}^{48}$ and $R_{\rm skin}^{208}$ may not be as strong as that observed between ${}^{208}$Pb and other 
heavier neutron-rich nuclei\,\cite{Piekarewicz:2012pp}, it is difficult to find a theoretical model than can reproduce both. 
Indeed, several theoretical approaches have  attempted unsuccessfully to reconcile both
measurements\,\cite{Hu:2021trw,Reinhard:2022inh,Zhang:2022bni,Mondal:2022cva,Papakonstantinou:2022gkt,Yuksel:2022umn}. The common 
theme that has emerged from these studies suggests that it is difficult to accommodate within $1\sigma$ the large value of $R_{\rm skin}^{208}$ within the constraints 
imposed from other nuclear observables, particularly the electric dipole polarizability $\alpha_D$.
Reference\,\cite{Reinhard:2022inh} goes as far as suggesting that the large error bars reported by the PREX collaboration makes it difficult to use the 
parity violating asymmetry in ${}^{208}$Pb as a meaningful constraint on the isovector sector of current energy density functionals. Finally, a recent ab 
initio approach using chiral forces predicted a value for $R_{\rm skin}^{208}$ that is in mild tension with PREX\,\cite{Hu:2021trw}, and for 
which the correlation between $R_{\rm skin}^{208}$ and $R_{\rm skin}^{48}$ remains fairly strong. Additionally, the range of parameter space inferred by ab initio calculations point to the large neutron skin of \lead being unlikely \cite{sym16010034}.
We note that within the ab initio framework, a prediction of the neutron skin thickness of ${}^{48}$Ca was made well in advance of the experiment\,\cite{Hagen:2015yea} that is entirely consistent with the value reported by the CREX collaboration. 
It is the goal of this work to explore refinements to the isovector sector of covariant energy density functionals in an attempt to reconcile the PREX-2 and 
CREX results. We first begin by showing the correlation of several energy density functional (EDF) predictions of isovector skins with some of the terms of the symmetry energy in \autoref{sec:correlations}. We then extract constraints on these terms inferred from the results of both PREX and CREX using a Markov Chain Monte Carlo (MCMC) approach in  \autoref{sec:bayesian}. Next, in \autoref{sec:dinoset}, we generate three new relativistic EDFs which are consistent with the results extracted above. By using these three new EDFs, we predict several properties of nuclear and neutron star matter (see \autoref{sec:predictions}). We then conclude in \autoref{sec:conclusions}.

\section{Isovector Skin Correlations}
\label{sec:correlations}
In a previous letter, we invoked the strong correlation between $R_{\rm skin}^{208}$ and $L$ to conclude that the symmetry energy is 
stiff\,\cite{Reed:2021nqk}. While it may be possible to soften the symmetry energy by incorporating the CREX results, such an approach
seems unnatural given that several recent studies have concluded that the current picture of density functional theory is insufficient to 
explain why ${}^{208}$Pb has a thick skin and ${}^{48}$Ca has a thin skin\,\cite{Reinhard:2022inh,Papakonstantinou:2022,Yuksel:2022umn}. 
What may hold the key to elucidate the source of the discrepancy are the higher order terms in the density expansion of the symmetry 
energy displayed in Eq.(\ref{eq:sym_energy}), particularly $\Ksym$. In what follows, we actually argue that the CREX results may provide 
the means to constrain $\Ksym$. This is relevant given that earlier studies have suggested that the properties of finite nuclei with a relative 
modest neutron-proton asymmetry are insensitive to $\Ksym$\,\cite{Tsang:2008fd,Tsang:2012se,Danielewicz:2016bgb}.

\begin{figure*}[htb]
\includegraphics[width=0.9\textwidth]{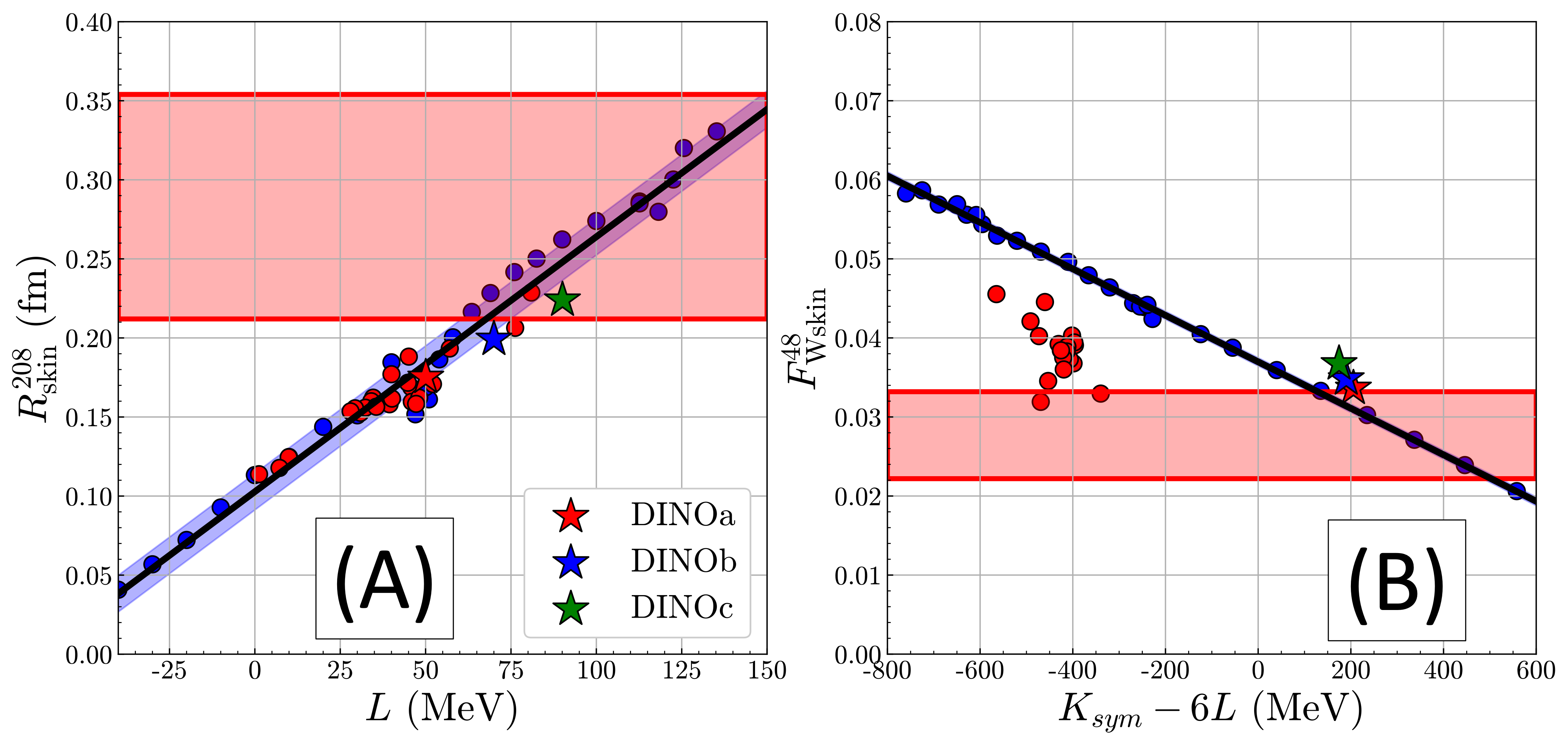}
\caption{(a) Model predictions for the neutron skin thickness of ${}^{208}$Pb as a function of $L$ and 
              (b) for the weak skin form factor of ${}^{48}$Ca as a function of the linear combination 
              $K_{\rm sym}\!-\!6L$. The blue and red dots represent the entire set of both covariant
               and non-relativistic EDFs used in the CREX analysis\,\cite{Adhikari:202kgg}. The new 
               DINOa-c interactions are shown as stars. The experimental values are shown as horizontal 
               bands and the $1\sigma$ errors bands for the covariant EDFs are displayed as solid lines with blue bands.}
\label{Fig1}
\end{figure*}

\begin{figure*}[htb]
\includegraphics[width=0.9\textwidth]{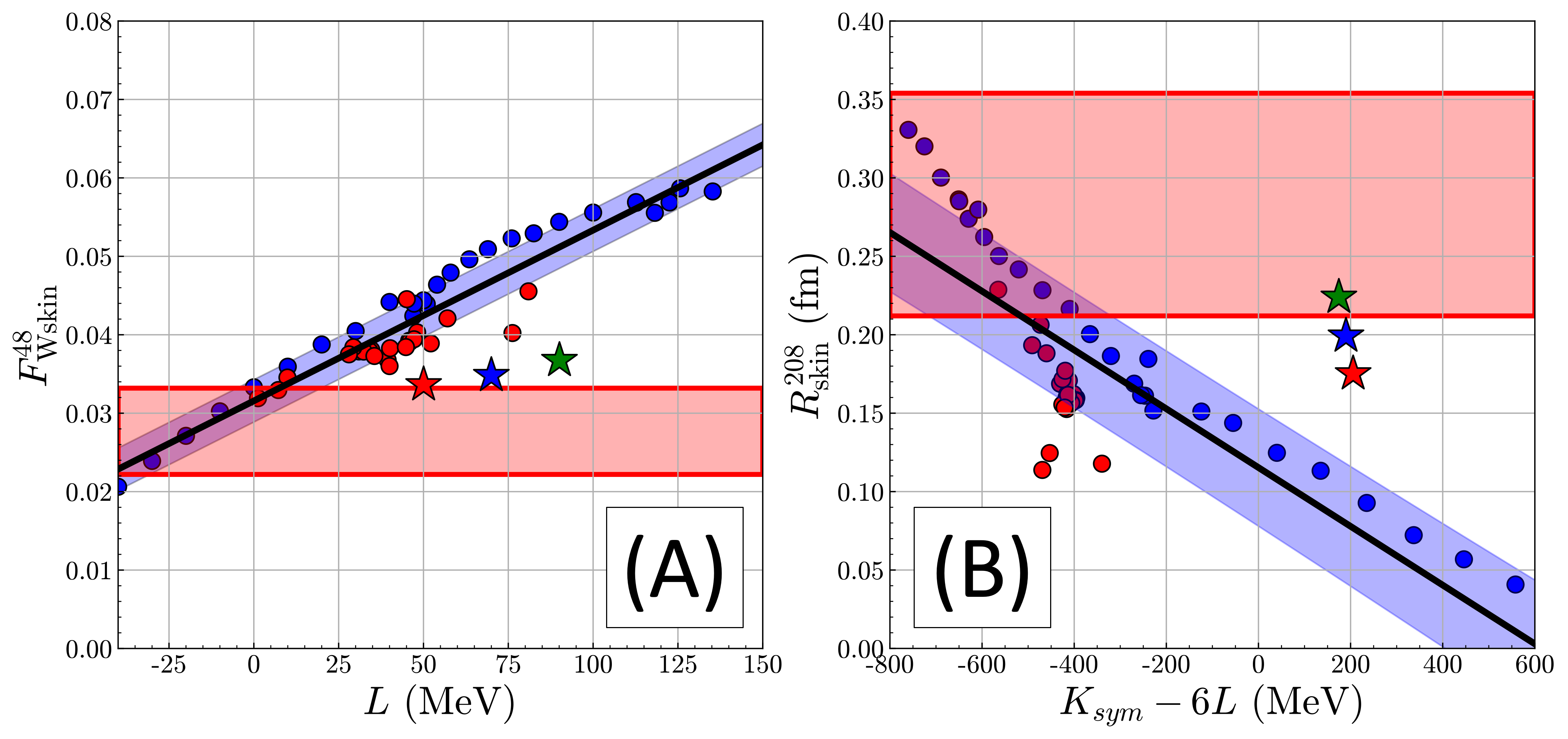}
\caption{(a) Model predictions for the weak skin form factor of ${}^{48}$Ca as a function of $L$ and 
              (b) for the neutron skin thickness of ${}^{208}$Pb as a function of the linear combination 
              $K_{\rm sym}\!-\!6L$. The blue and red dots represent the entire set of both covariant
               and non-relativistic EDFs used in the CREX analysis\,\cite{Adhikari:202kgg}. The new 
               DINOa-c interactions are shown as stars, see \autoref{Fig1}. The experimental values are shown as horizontal 
               bands and our linear fits with $1\sigma$ errors are shown as solid lines with a blue band.}
\label{Fig2}
\end{figure*}

We begin by displaying in Fig.\ref{Fig1}a the well-known correlation between $L$ and the neutron skin thickness of ${}^{208}$Pb. 
For this work, we employ the entire set of models used in the CREX analysis, which includes a fairly comprehensive set of covariant 
energy density functionals (blue dots) and a limited set of non-relativistic Skyrme models (red dots). Predictions are also included for 
three new covariant EDFs labeled as ``DINOa-c" (colored stars) that are introduced below and in Table\,\ref{tab:BE_RC}. In both Fig.\ref{Fig1} and 
Fig.\ref{Fig2} we display 1$\sigma$ theoretical error bands obtained by including only the covariant EDFs displayed by the blue dots. Although 
the correlation between $L$ and the neutron skin thickness of ${}^{48}$Ca was anticipated not to be as robust as in the case of 
${}^{208}$Pb, Fig.\ref{Fig1}(b) suggests that for the entire set of covariant EDFs, a very strong correlation emerges between 
$K_{\rm sym}\!-\!6L$ and the CREX weak skin form factor $F_{\rm Wskin}^{48}$ evaluated at the average momentum transfer of the experiment 
$q\!=\!0.8733\,{\rm fm}^{-1}$.

This figure suggests---for the set of covariant EDFs employed in this work---the CREX form factor may best be correlated 
with linear combinations of the symmetry energy parameters. In particular, the quantity $\Ksym-6L$ appears in the incompressibility coefficient of asymmetric matter $K(\alpha) = K_{0} + K_\tau\,\alpha^{2} + \ldots$, where 
\begin{eqnarray}
    K_\tau = K_{\mathrm{sym}}-6L-\frac{Q_0}{K_0}L,
    \label{eq:ktau}
\end{eqnarray}
Here $\alpha\!=\!(N\!-\!Z)/A$, $K_{0}$ is the incompressibility coefficient of symmetric nuclear matter, and $Q_0$ the
corresponding skewness parameter\,\cite{Piekarewicz:2008nh}. Note that although motivated by the incompressibility 
of asymmetric matter, we find that $F_{\rm Wskin}^{48}$ is slightly better correlated to $K_{\rm sym}\!-6L$ than to 
$K_\tau$ because of the small model dependency coming from the third term. We also note that whereas the correlation with the covariant EDFs is very sharp, the non-relativistic models 
do not follow the same trend. However, the collection of non-relativistic EDFs used here is not representative, as
it spans a fairly narrow range in $K_{\rm sym}\!-\!6L$. Hence, investigating whether this correlation develops as one 
includes a comprehensive set of non-relativistic models is warranted.

As already mentioned and now displayed in Fig.\ref{Fig2}, neither the correlation between $F_{\rm Wskin}^{48}$ and $L$ 
nor the correlation between $R_{\rm skin}^{208}$ and $K_{\rm sym}\!-\!6L$ is as strong as those displayed in Fig.\ref{Fig1}. Nevertheless, we show this behavior to clarify some of the details of the statistical inference that will be implemented later on.

\section{Bayesian Inference of L and $\Ksym$}
\label{sec:bayesian}

Although we caution against the universality of the correlation between $\Fskin{48}$ and $\Ksym\!-6L$, we nevertheless 
proceed to constrain both $L$ and $\Ksym$ using the PREX-2 and CREX results. To account for a possible model dependence, 
we consider two different scenarios. In Scenario 1 we determine $L$ from PREX-2 
and then use the newly-found correlation between $F_{\rm Wskin}^{48}$ and $\Ksym-6L$ to extract $\Ksym$. In this scenario 
only the set of covariant EDFs is used together with the correlations identified in Fig.\ref{Fig1}. In Scenario 2 we now add the limited set of Skyrme forces together with the correlations displayed in Fig.\ref{Fig2}. Under
this scenario, which greatly increases the model uncertainty, the 1$\sigma$ theoretical error bands are now expanded 
to include both non-relativistic and covariant EDFs. We underscore that such analysis should 
be repeated using a larger set of Skyrme EDFs, given that the limited set included here spans a narrow range of values 
for $\Ksym\!-6L$.

For our calculation of optimal $L$ and $\Ksym$ values, we employ the use of a $\chi^2$ minimization routine. The $\chi^2$ takes into account the experimental values and the linear fits shown in \autoref{Fig1} and \autoref{Fig2}. For the experimental errors, we use the results of the main PREX-2 \cite{Adhikari:2021phr} and CREX \cite{Adhikari:202kgg} papers. We construct our $\chi^2$ using the standard form:
\begin{equation}
    \chi^2 = \sum \frac{\Big(y_i-\hat{y_i}(L,K_{\rm sym})\Big)^2}{\sigma_{\rm th}^2+\sigma_{\rm exp}^2}
\end{equation}
which we minimize using downhill simplex method \cite{scipy}. Here $y_i$ is a set of experimental observables, whose theoretical predictions $\hat{y}_i(L,K_{\rm sym})$ are sensitive to the symmetry energy parameters $L$, and $K_{\rm sym}$, whereas $\sigma_{\rm th}$ and $\sigma_{\rm exp}$ indicate $1\sigma$ theoretical and experimental errors, respectively. 

The list of models we employ in calculating our linear correlations come directly from the main CREX paper \cite{Adhikari:202kgg}, given in \autoref{tab:models1} and \autoref{tab:models2}. Important for our determination of both parameters is the model uncertainty $\sigma_{\rm th}$ used in the $\chi^2$. To do this, we calculate the 1$\sigma$ prediction interval (PI) from the linear regression line. At each point along the regression line, the $1\sigma$ PI is calculated using
\begin{eqnarray}
    \sigma_{\rm th} = t_{68\%}\times \sqrt{\frac{\sum_{i=1}^{n}(y_i-\hat{y}(x_i))^2}{n-2}}
\end{eqnarray}
where $t_{68\%}$ is the critical t-value at 68\% confidence and the terms in the square-root represent the standard deviation of the residuals. The prediction interval gives us an approximate spread of values around the regression line which indicates the region where $\sim68\%$ of all models should fall. For the Scenario 1 fit, we fit the regression line to the covariant models only for the relations in \autoref{Fig1}. For the Scenario 2 fit, we use the regression line fit to both covariant and the Skyrme models.

\begin{table}[htb]
    \centering
    \begin{tabular}{l|c|c|c|c}\hline\hline
Model & $L$ & $\Ksym$ & $\Fskin{48}$ & $\Rskin{208}$ \\
& (MeV) & (MeV) & & (fm)\\\hline
FSUGarnet \cite{Chen:2015zpa} & 50.97 & 59.45 & 0.0439 & 0.1611 \\
FSUGold2 \cite{Chen:2014mza} & 112.68 & 25.38 & 0.0568 & 0.2863 \\
FSUGold2 (47) \cite{Fattoyev:2017jql} & 47.0 & 53.98 & 0.0424 & 0.1517 \\
FSUGold2 (50) \cite{Fattoyev:2017jql} & 50.0 & 30.23 & 0.0444 & 0.1688 \\
FSUGold2 (54) \cite{Fattoyev:2017jql} & 54.0 & 3.67 & 0.0464 & 0.1864 \\
FSUGold2 (58) \cite{Fattoyev:2017jql} & 58.0 & -17.8 & 0.0479 & 0.2004 \\
FSUGold2 (69) \cite{Fattoyev:2017jql} & 69.0 & -54.97 & 0.0509 & 0.2284 \\
FSUGold2 (76) \cite{Fattoyev:2017jql} & 76.0 & -64.40 & 0.0523 & 0.2416 \\
FSUGold2 (90) \cite{Fattoyev:2017jql} & 90.0 & -55.47 & 0.0544 & 0.2623 \\
FSUGold2 (100) \cite{Fattoyev:2017jql} & 100.0 & -29.14 & 0.0556 & 0.2740 \\
IUFSU \cite{Fattoyev:2010mx} & 47.21 & 28.53 & 0.044 & 0.1615 \\
NL3 \cite{Lalazissis:1996rd} & 118.2 & 100.88 & 0.0555 & 0.2798 \\
RMF022 \cite{Chen:2015zpa} & 63.52 & -28.76 & 0.0496 & 0.2164 \\
RMF028 \cite{Chen:2015zpa} & 112.65 & 26.26 & 0.0569 & 0.2851 \\
RMF032 \cite{Chen:2015zpa} & 125.63 & 28.68 & 0.0587 & 0.3201 \\
TAMUa \cite{Fattoyev:2013yaa} & 82.49 & -68.37 & 0.0529 & 0.2502 \\
TAMUb \cite{Fattoyev:2013yaa} & 122.53 & 45.88 & 0.0569 & 0.3002 \\
TAMUc \cite{Fattoyev:2013yaa} & 135.25 & 51.64 & 0.0583 & 0.3306 \\
IU$\delta$ (-40) \cite{Adhikari:202kgg} & -40.0 & 318 & 0.0206 & 0.0408 \\
IU$\delta$ (-30) \cite{Adhikari:202kgg} & -30.0 & 266 & 0.0239 & 0.0568 \\
IU$\delta$ (-20) \cite{Adhikari:202kgg} & -20.0 & 217 & 0.0271 & 0.0722 \\
IU$\delta$ (-10) \cite{Adhikari:202kgg} & -10.0 & 175 & 0.0302 & 0.0928 \\
IU$\delta$ (0) \cite{Adhikari:202kgg} & 0.0 & 135 & 0.0333 & 0.1133 \\
IU$\delta$ (10) \cite{Adhikari:202kgg} & 10.0 & 100 & 0.0359 & 0.1247 \\
IU$\delta$ (20) \cite{Adhikari:202kgg} & 20.0 & 65 & 0.0388 & 0.1438 \\
IU$\delta$ (30) \cite{Adhikari:202kgg} & 30.0 & 56 & 0.0405 & 0.1510 \\
IU$\delta$ (40) \cite{Adhikari:202kgg} & 40.0 & 1.0 & 0.0442 & 0.1846 \\
\end{tabular}
\caption{List of all relativistic energy density functional (EDF) models used in the CREX analysis. Note there are several FSUGold2 and IU-$\delta$ models with different values of the slope parameter of the symmetry energy $L$. Also listed are $\Rskin{208}=R_n-R_p$ for \lead and $\Fskin{48}=F_{ch}-F_{wk}$ for $^{48}$Ca.}
    \label{tab:models1}
\end{table}

\begin{table}[htb]
\caption{List of all non-relativistic EDF models used in the CREX analysis, similar to \autoref{tab:models1}.}
    \centering
    \begin{tabular}{l|c|c|c|c}\hline\hline
Model & $L$ & $\Ksym$ & $\Fskin{48}$ & $\Rskin{208}$ \\
& (MeV) & (MeV) & & (fm)\\\hline
SI \cite{Skyrme:1956zz} & 1.22 & -461.85 & 0.0319 & 0.1138 \\
SIII \cite{Vautherin:1971aw} & 9.91 & -393.74 & 0.0345 & 0.1246 \\
SKM* \cite{BARTEL198279} & 45.78 & -155.94 & 0.0392 & 0.1688 \\
SLy4 \cite{CHABANAT1998231} & 45.96 & -119.7 & 0.0391 & 0.1596 \\
SLy5 \cite{reinhard:1999} & 48.14 & -112.76 & 0.0403 & 0.1622 \\
SLy7 \cite{reinhard:1999} & 47.22 & -113.32 & 0.0394 & 0.1583 \\
SV-K218 \cite{Klupfel:2008af} & 34.62 & -206.88 & 0.0381 & 0.1622 \\
SV-K226 \cite{Klupfel:2008af} & 34.09 & -211.92 & 0.0379 & 0.1599 \\
SV-K241 \cite{Klupfel:2008af} & 30.95 & -230.77 & 0.0379 & 0.1527 \\
SV-bas \cite{Klupfel:2008af} & 32.36 & -221.76 & 0.0379 & 0.1559 \\
SV-kap00 \cite{Klupfel:2008af} & 39.44 & -161.78 & 0.0368 & 0.158 \\
SV-kap02 \cite{Klupfel:2008af} & 35.54 & -193.2 & 0.0373 & 0.1565 \\
SV-kap06 \cite{Klupfel:2008af} & 29.33 & -249.76 & 0.0384 & 0.1555 \\
SV-mas07 \cite{Klupfel:2008af} & 52.15 & -98.77 & 0.0389 & 0.1708 \\
SV-mas08 \cite{Klupfel:2008af} & 40.15 & -172.39 & 0.0383 & 0.1616 \\
SV-mas10 \cite{Klupfel:2008af} & 28.03 & -252.51 & 0.0375 & 0.1536 \\
SV-sym28 \cite{Klupfel:2008af} & 7.21 & -296.51 & 0.033 & 0.1178 \\
SV-sym32 \cite{Klupfel:2008af} & 57.07 & -148.8 & 0.0421 & 0.1933 \\
SV-sym34 \cite{Klupfel:2008af} & 80.95 & -79.08 & 0.0455 & 0.2287 \\
SV-min \cite{Klupfel:2008af} & 44.81 & -156.57 & 0.0384 & 0.1716 \\
TOV-min \cite{Erler:2012qd} & 76.23 & -15.62 & 0.0402 & 0.2064 \\
UNEDF0 \cite{Kortelainen:2010dt} & 45.08 & -189.68 & 0.0445 & 0.1882 \\
UNEDF1 \cite{Kortelainen:2010hv} & 40.00 & -179.48 & 0.036 & 0.177 \\
\end{tabular}
    \label{tab:models2}
\end{table}

\begin{figure}[h]
 \centering
 \includegraphics[width=\columnwidth]{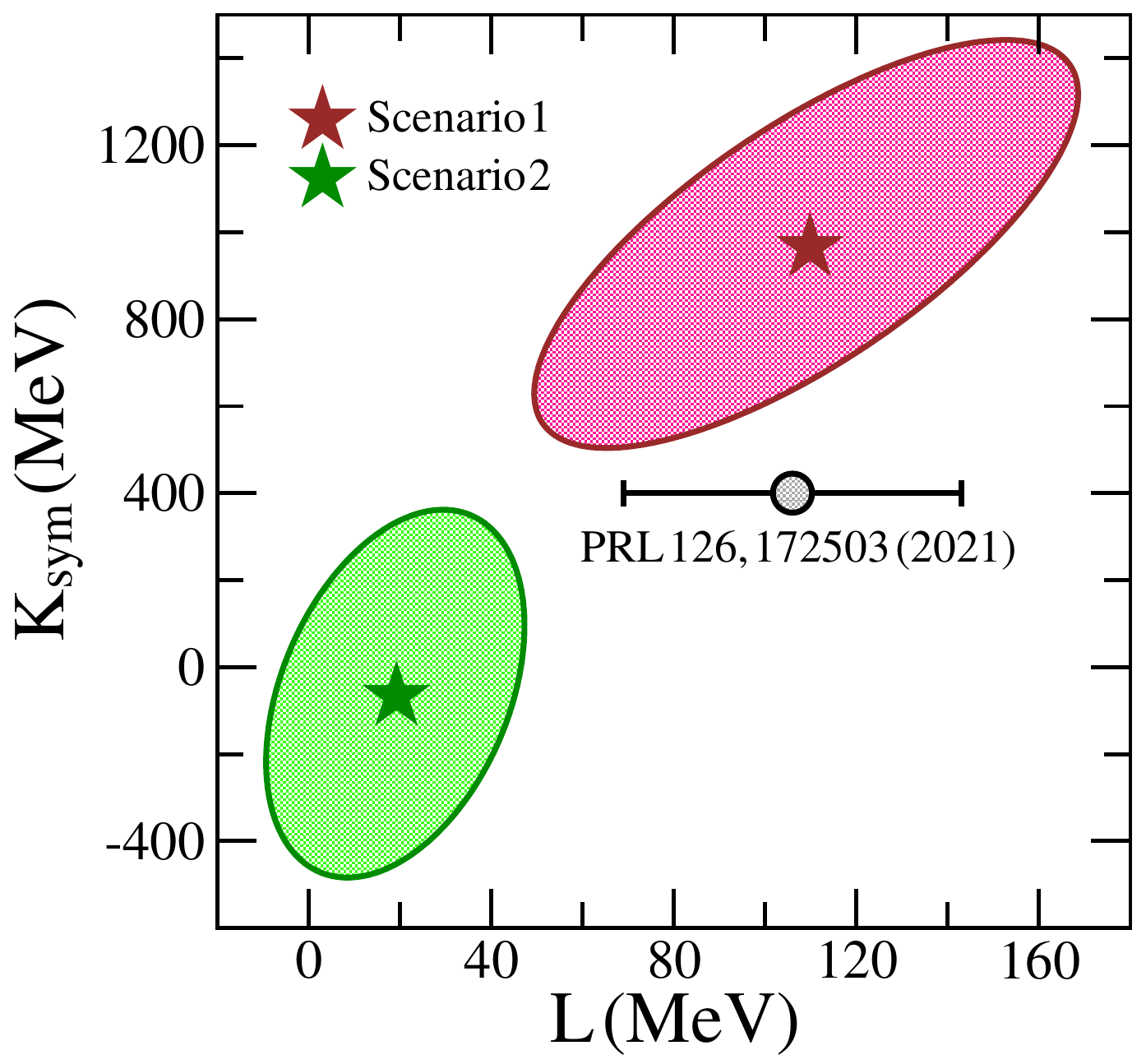}
 \caption{Confidence ellipses for $L$ and $\Ksym$ from \autoref{Table1}. We show the central values of each fit as a star with the surrounding 
               ellipse indicating the 67\% confidence interval. Notably, the Scenario 1 and Scenario 2 fits do not overlap.}
 \label{fig:ellipses}
\end{figure}

Once optimal values are found from the simplex method, the posterior distribution is found by maximizing the likelihood function, defined as $\mathcal{L}=\exp{(-\frac{1}{2}\chi^2)}$, using a Markov Chain Monte Carlo approach \cite{emcee}. We use a uniform prior on this approach for $L$ in the range of $0<L<200$ MeV and leave $\Ksym$ unconstrained. Of importance here is to note that the model error is interpolated over the x-coordinate by taking half the difference of the upper and lower PI bound at each point along the regression line. Once the posterior distribution is sampled, one obtains a distribution of model parameters from which averages and standard deviations may be computed. Our results for both scenarios can be found in \autoref{Table1} with confidence ellipses shown in \autoref{fig:ellipses}.

 \begin{table}[htb]
    \begin{tabular}{|c | c | c | c|}\hline
        Fit & L (MeV) & $\Ksym$ (MeV)\\\hline
        Scenario 1 & $110\pm40$ & $970\pm320$\\
        Scenario 2 & $19\pm19$ & $-61\pm280$\\ \hline
   \end{tabular}
   \caption{Averages and standard deviations for the two scenarios
                 discussed in the text.}
   \label{Table1}
  \end{table}

Our results paint an interesting picture of the density dependence of the symmetry energy. First, the value of $L\!=\!110\pm40$\,MeV 
extracted from Scenario 1 is entirely consistent with our original published value of $L\!=\!106\pm37$\,MeV\,\cite{Reed:2021nqk},
that is also displayed in the figure. The small differences are attributed to the larger set of covariant EDFs used here. In contrast, Scenario 2 favors a considerably smaller value of $L\!\approx\!19\pm19$\,MeV. Such a smaller value is driven by the small experimental error in $F_{\rm Wskin}^{48}$ relative to $F_{\rm Wskin}^{208}$. So even if the correlation to $L$ is
weaker for the former than for the latter, the smaller experimental error generates the significantly smaller value of $L$.

Whereas our knowledge of $L$ has improved since the culmination of the PREX campaigns, $\Ksym$ has remained 
largely unconstrained, with an overwhelming number of theoretical approaches favoring negative values. In contrast, 
Scenario 1 favors positive values for $\Ksym$ at the $3\sigma$ level. It is only by including the limited set of 
non-relativistic models, in combination with the small values of $L$, that $\Ksym$ is allowed to take negative values. 
We find this possible conflict particularly interesting. After all, one of the main motivations behind CREX was the use 
of ab initio models to inform and improve the isovector sector of energy density functionals\,\cite{CREX:2013}. 

\section{Newly Created Covariant Model Set}
\label{sec:dinoset}

However, before incorporating ab initio predictions, we adopt the Scenario 1 PREX-2--CREX constraints in an attempt to generate
three new covariant EDFs with large and positive values for $\Ksym$. These models, named ``DINOa,b,c", are members 
of the ``FSUGold'' class of covariant EDFs\,\cite{Horowitz:2001ya,Todd-Rutel:2005fa}, but with a notable addition to the 
isovector sector. In the past, the isovector sector of this class of models included only two terms: a Yukawa coupling of the
nucleon to the isovector $\rho$-meson and a non-linear $\omega$-$\rho$ isoscalar-isovector cross-coupling term. In this 
work we add a contribution to the isovector sector arising from the Yukawa coupling of the nucleon to the scalar-isovector 
$\delta$-meson\,\cite{Kubis:1998jt, Hofmann:2000vz, Liu:2001iz, Greco:2002sp, Bunta:2003fm, Liu:2004yt}. The presence 
of the $\delta$-meson splits the nucleon effective mass, thereby providing an additional degree of freedom in selecting 
$\Ksym$.  To span the range of reasonable values for $L$, we have defined the DINOa, DINOb, and DINOc, models by
fixing the slope of the symmetry energy at $L\!=\!50,70,90$\,MeV, respectively. Our fitting procedure follows closely that of 
Ref.\,\cite{Fattoyev:2010mx}, supplemented by the inclusion of both $R_{\rm skin}^{208}$ and $\Fskin{48}$ into the 
calibration. 

We begin our construction with the class of FSUGold-like models featuring the delta meson \cite{Todd-Rutel:2005fa,Singh:2014}. The interaction Lagrangian for this class of models is
\begin{eqnarray}
    \mathcal{L}_{\mathrm{int}}&&=\bar{\psi}\Big[\mathcal{S}(\phi,\bm{\delta})-\mathcal{V}_\mu(V_\mu,\bm{b_\mu},A_\mu)\gamma^\mu\Big]\psi-U(\phi)\\\nonumber
    &&+\frac{\zeta}{4!}g_v^4(V_\mu V^\mu)^2+\Lambda_vg_v^2g_\rho^2V_\mu V^\mu\bm{b_\mu}\cdot\bm{b^\mu}
    \label{eq:lagrangian}
\end{eqnarray}
where
\begin{eqnarray}
    &&\mathcal{S}(\phi,\bm{\delta}) = g_s\phi+\frac{g_\delta}{2}\bm{\tau}\cdot\bm{\delta}\\\nonumber
    &&\mathcal{V}_\mu(V_\mu,\bm{b_\mu},A_\mu) = g_vV_\mu+\frac{g_\rho}{2}\bm{\tau}\cdot \bm{b_\mu}+\frac{e}{2}(1+\tau_3)A_\mu\\\nonumber
    &&U(\phi) = \frac{\kappa}{3!}(g_s\phi)^3+\frac{\lambda}{4!}(g_s\phi)^4
\end{eqnarray}
This class of models contain the interactions of nucleons ($\psi$) via the exchange of the scalar $\sigma$ ($\phi$) and vector $\omega$ ($V_\mu$) mesons as well as the vector-isovector $\rho$-meson ($\textbf{b}^\mu$), and, in this work, the additional vector-isoscalar $\delta$-meson ($\bm{\delta}$). In addition to interactions among the nucleons, this Lagrangian class also includes many self and cross-coupling terms. These are $\kappa$ and $\lambda$ for the cubic and quartic scalar self-coupling, $\zeta$ for the quartic $\omega$ self-coupling, and $\Lambda_v$ as the $\omega$-$\rho$ cross coupling.

Motivated by our findings of large positive $\Ksym$ value from the Scenario 1 fit, we choose to generate three models which we expect run the gamut on parameter space. The generation of new models follows a similar prescription as described in \cite{Fattoyev:2010rx} which we summarize here briefly. For each model, we fix the value of the symmetry energy at sub-saturation density $k_F=1.15$ fm$^{-1}$ to be $\tilde{J}=27$ MeV. This value has been shown to agree well with measurements of nuclear masses. We then fix $L$ for the three models at 50, 70, and 90 MeV to give a good range of $L$ values favored from multiple sources including PREX-2. We also fix the $\zeta$-parameter to $0.015$, which controls the high-density components of the EOS. This allows our models to predict neutron star masses consistent with the most recent observational and theoretical constraints on the maximum mass\,\cite{Demorest:2010bx, Antoniadis:2013pzd, Rezzolla:2017aly, Miller:2021qha, Riley:2021pdl, Romani:2022jhd}. From here, we take a typical set of bulk nuclear matter properties, and adjust the isoscalar bulk properties to fit the binding energy and charge radii (with spin-orbit corrections \cite{Horowitz:2009ya}) of \ca, \calfe, and \lead. 

The inversion of bulk properties to RMF Lagrangian couplings has been described in \cite{Chen:2014mza}. It was found that the Giant Monopole Resonances (GMR) are sensitive primarily to the incompressibility of neutron-rich matter, $K(\alpha)$\,\cite{Piekarewicz:2009gb}. In our previous work\,\cite{Fattoyev:2013yaa} we found that for $^{208}$Pb this corresponds to $K(\alpha) \approx 222$ MeV for the RMF parameterizations that yield consistent GMRs. Hence we applied this constraint to $K(\alpha)$ in the DINO models.  All that is left now is to pick a value for $\Ksym$ which minimizes the squared difference of $\Rskin{208}$ and $\Fskin{48}$ from their experimental values. The difference between the model predictions and experimental results for each observable are then used to generate a new set of parameters until a parameter set converges.

It should be noted that a direct inversion of couplings from $\tilde{J}$, $L$, and $\Ksym$ does not presently exist as the presence of the additional $\delta$-meson introduces much more complexity than has previously been detailed. As such, we use a downhill simplex routine to determine the isovector couplings $g_\rho^2$, $g_\delta^2$, and $\Lambda_v$. Furthermore, there exist some combinations of bulk nuclear properties that do not produce a convergent set of parameters which requires careful calculation schemes.

The coupling constants of the three DINO models can be found in \autoref{tab:couplings}. We find that due to the large value of $\Ksym$, the isovector Yukawa couplings $g_\rho^2$ and $g_\delta^2$ are very large. Additionally, we report the bulk parameters of the nuclear EOS in \autoref{tab:bulk_props} as predicted by these three models. In the next section, using these new parameters set we proceed to calculate the properties of finite nuclei and neutron stars.

\begin{table*}[htb]
    \centering
    \begin{tabular}{|c|c|c|c|c|c|c|c|c|c|c|c|c|}
    \hline
       Model & $m_{\rm s}$ &  $g_s^2$ & $g_{\delta}^2$ & $g_v^2$ & $g_{\rho}^2$ & $\kappa$ & $\lambda$ & $\zeta$ & $\Lambda_{\rm v}$ \\
    \hline
       DINOa & $490.050$ & $93.9422$ & $1115.15$ & $154.436$ & $805.891$ & $4.9860$ & $-0.01370$ & $0.015$ & $0.0016497$ \\
    \hline
       DINOb & $485.795$ & $91.0316$ & $1252.71$ & $150.824$ & $877.121$ & $5.2914$ & $-0.01488$ & $0.015$ & $0.0014014$ \\ 
    \hline
       DINOc & $484.162$ & $90.6481$ & $1343.25$ & $151.048$ & $922.617$ & $5.3209$ & $-0.01497$ & $0.015$ & $0.0012312$ \\
    \hline                
    \end{tabular}
    \caption{Coupling constants of the DINO set of models. For each of the three models, we fix $m_{\rm v}=782.5$ MeV, $m_{\rho}=763$ MeV, and $m_\delta=980$ MeV as well as the quartic vector self-coupling $\zeta=0.015$. Scalar meson mass and nonlinear $\kappa$ values given in MeV.}
    \label{tab:couplings}
\end{table*}

\begin{table*}[htb]
    \centering
    \begin{tabular}{|c|c|c|c|c|c|c|c|c|c|c|}
    \hline
       Model & $\rho_0$ (fm$^{-3}$) &  $\epsilon_0$ (MeV) & $K_0$ (MeV) & $Q_0$ (MeV) & $\tilde{J}$ (MeV) & $J$ (MeV) & $L$ (MeV) & $K_{\rm sym}$ (MeV) & $K_{\tau}$ (MeV) & $K(\alpha)$ (MeV)  \\
    \hline
       DINOa & $0.1522$ & $-16.16$ & $210.0$ & $-361.4$ & $27.00$ & $31.42$ & $50.00$ & $506.0$ & $292.1$ & $223.1$ \\
    \hline
       DINOb & $0.1525$ & $-16.21$ & $207.0$ & $-412.0$ & $27.00$ & $33.07$ & $70.00$ & $610.0$ & $329.3$ & $221.7$ \\ 
    \hline
       DINOc & $0.1519$ & $-16.22$ & $206.0$ & $-426.1$ & $27.00$ & $34.58$ & $90.00$ & $715.0$ & $361.2$ & $222.2$ \\ 
    \hline                
    \end{tabular}
    \caption{Bulk properties of nuclear matter for our set of DINO models. Tilde denotes symmetry energy values taken at subsaturation density ($\tilde{k_F}=1.15$fm$^{-1}$). Here $K(\alpha)$ is calculated using the equation 18c of \cite{Piekarewicz:2008nh} for $\alpha = (126-82)/208$.}
    \label{tab:bulk_props}
\end{table*}

\begin{figure}
    \centering
    \includegraphics[width=\columnwidth]{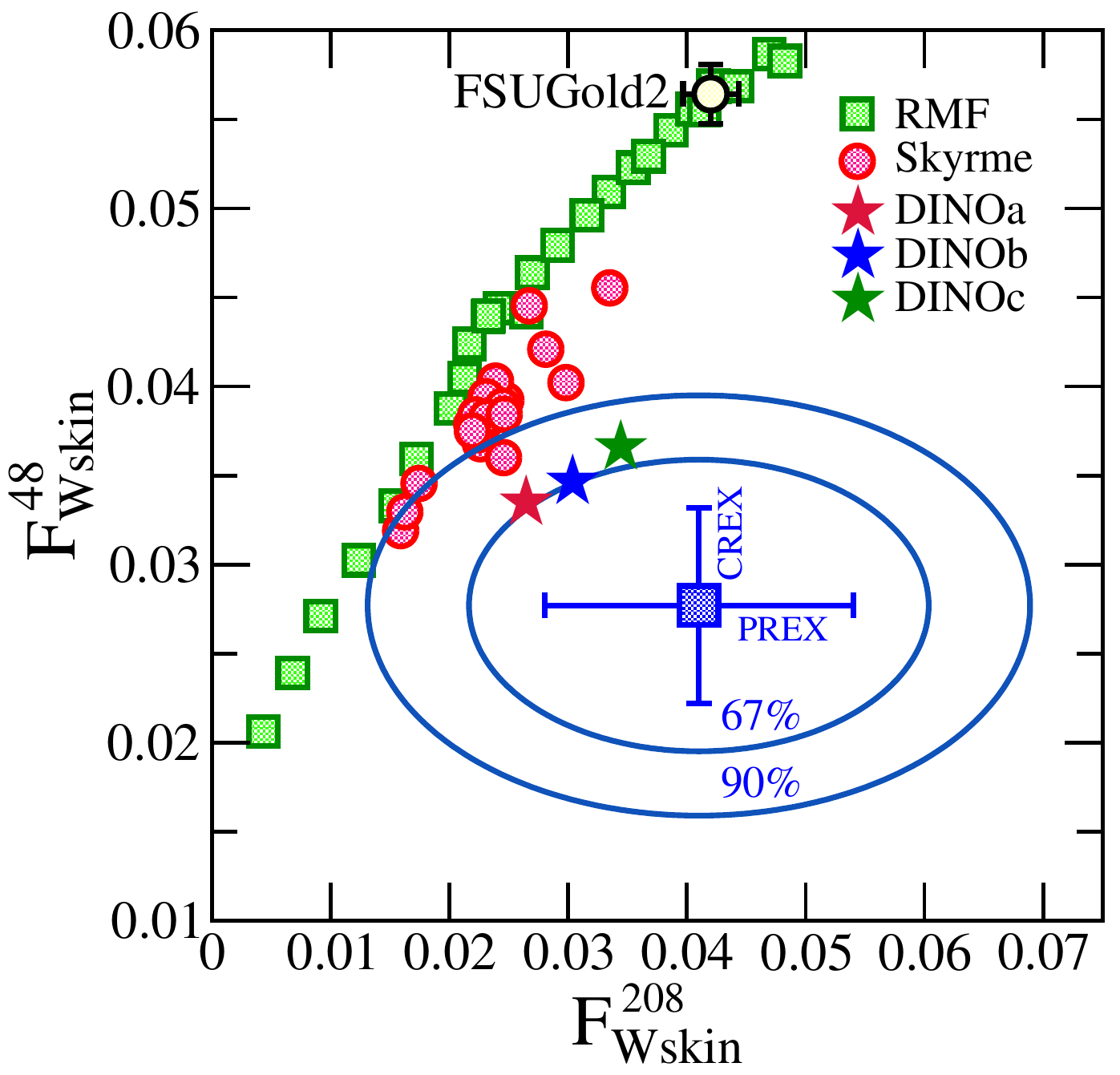}
    \caption{Predictions for the weak skin form factor of ${}^{208}$Pb and ${}^{48}$Ca 
    for the entire set of covariant RMF (green squares) and non-relativistic Skyrme (pink circles) EDFs considered in this letter. The blue ellipses 
    represent joint PREX-2 and CREX 67\% and 90\% probability contours, just as in Figure 2 of Ref.\cite{Adhikari:202kgg}. 
    The FSUGold2 prediction is included to illustrate typical statistical uncertainties. Depicted with stars are the prediction 
    of the three new DINO models.}
  \label{fig:ffacs}
\end{figure}

\begin{table*}[htb]
    \centering
    \begin{tabular}{|c|c|c|c|c|c|c|c|c|}
    \hline
       Nucleus & Observable & Experiment & IUFSU & FSUGold2 & BigApple & DINOa & DINOb & DINOc \\
    \hline
       & $B/A$ (MeV) & 8.667 & 8.549 & 8.621 & 8.531 & 8.670 & 8.667 & 8.667 \\
                & $R_{\rm ch}$ (fm) & $3.477\pm0.002$ & 3.416 & 3.413 & 3.447 & 3.454 & 3.458 & 3.461 \\
    $^{48}$Ca            & $R_{\rm skin}$ (fm) & $0.121 \pm 0.035$ & 0.1731 & 0.2319 & 0.1682 & 0.0996 & 0.1051 & 0.1135 \\
        & $\Fskin{}(q_{\mathrm{CREX}})$ & $0.0277\pm 0.0055$ & 0.0439 & 0.0568 & 0.0417 & 0.0333 & 0.0349 & 0.0367 \\  
    \hline
       & $B/A$ (MeV) & 7.867 & 7.896 & 7.872 & 7.872 & 7.867 & 7.867 & 7.867 \\
                & $R_{\rm ch}$ (fm) & $5.501\pm0.001$ & 5.476 & 5.489 & 5.490 & 5.502 & 5.503 & 5.503 \\
    $^{208}$Pb & $R_{\rm skin}$ (fm) & $0.283 \pm 0.071$ & 0.1615 & 0.2863 & 0.1506 & 0.1748 & 0.1993 & 0.2240 \\
                & $\Fskin{}(q_{\mathrm{PREX}})$ & $0.041\pm0.013$ & 0.0233 & 0.0423 & 0.0215 & 0.0263 & 0.0303 & 0.0344 \\
    \hline                
    \end{tabular}
    \caption{Experimental data for the binding energy per nucleon (in MeV)\,\cite{Wang:2012}, charge radii (in fm)\,\cite{Angeli:2013}, neutron skins, form factor skins\,\cite{Adhikari:2021phr, Adhikari:202kgg} for $^{48}$Ca and $^{208}$Pb nuclei used in the optimization. Also displayed are the theoretical results obtained with IUFSU\,\cite{Fattoyev:2010mx}, FSUGold2\,\cite{Chen:2014sca}, BigApple\,\cite{Fattoyev:2020cws}, and the three new parameterizations.}
    \label{tab:BE_RC}
\end{table*}

\section{DINO Model Predictions}
\label{sec:predictions}
\subsection{Predictions of Nuclei}
We start this section by mentioning that the new model parameters, together with predictions for the bulk properties of infinite nuclear matter are given \autoref{tab:couplings} and \autoref{tab:bulk_props}, respectively. It is worth mentioning, however, that in all three cases, the curvature of the symmetry energy is 
large and positive, namely, $K_{\rm sym}\!=\!506, 610, 715$ MeV, for DINOa, DINOb, and DINOc, respectively. Regardless, we find that the results for the binding energy, charge radii, and form factors of ${}^{48}$Ca and ${}^{208}$Pb as listed in Table\,\ref{tab:BE_RC} are all in good agreement with the experiment. In the particular case of $\Fskin{208}$ and $\Fskin{48}$, predictions for the large 
set of models considered here alongside the predictions from the three DINO models are displayed in Fig.\ref{fig:ffacs}. This figure highlights the predicament faced by most energy density functionals: with few exceptions, most models fall outside 
the 90\% confidence ellipse. In this sense, the three DINO models do quite well compared to the other models considered.

However, the DINO models' large isovector couplings do begin to introduce anomalies in the point particle densities. In particular, the proton and neutron densities in the core of both \calfe and \lead show large density fluctuations that are non-physical. For example, \lead is a heavy closed-shell nucleus and so we expect the nucleus to saturate at a density value near $\rho_0$ \cite{DeJager:1987qc,Horowitz:2020evx}. For the DINO models, \lead has very large fluctuations and \textit{does not} saturate. This anomaly is largely due to the isovector $g_\delta^2$ coupling splitting the proton and neutron masses, thereby causing a larger proton density in the core and a small neutron density. We see similar fluctuations in \calfe, although the saturation property is less strict for lighter nuclei.

\begin{figure}[htb]
    \centering
    \includegraphics[width=\columnwidth]{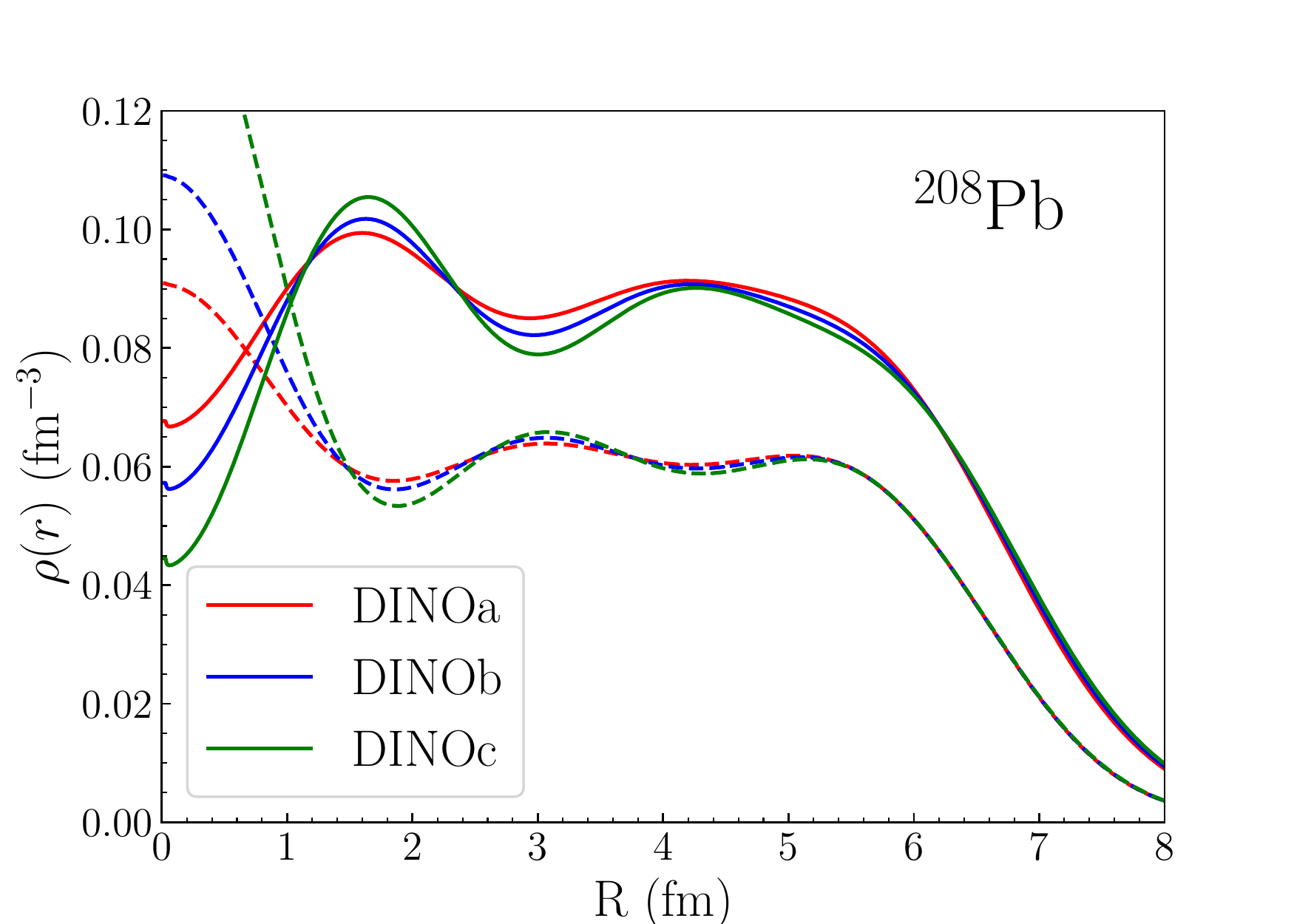}
    \caption{Point neutron (solid) and proton (dashed) densities in \lead for the DINO family of models. Note that the central proton density for DINOc is shown off the plot.}
    \label{fig:pb-dino-dens}
\end{figure}

\begin{figure}[htb]
    \centering
    \includegraphics[width=\columnwidth]{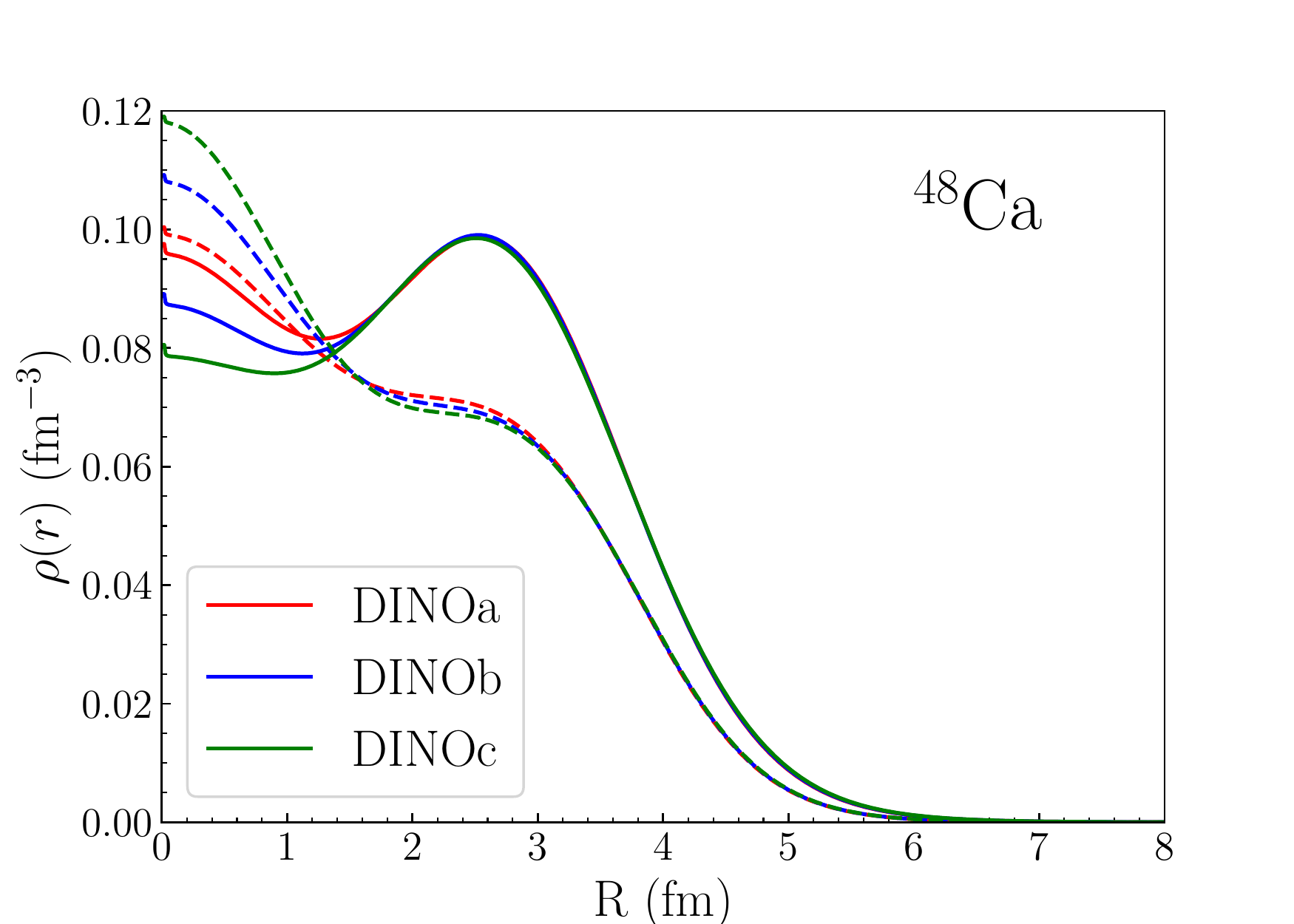}
    \caption{Point neutron (solid) and proton (dashed) densities in \calfe for the DINO family of models.}
    \label{fig:ca-dino-dens}
\end{figure}

\subsection{Predictions of Neutron Star Properties}
Our calculation of neutron star (NS) structure begins by following the standard practices of relativistic mean field theory to calculate the equation of state of the core \cite{Glendenning:2000,Fattoyev:2011,Singh:2014}. This includes calculating the pressure and energy density of protons, neutrons, electrons, and muons in beta equilibrium using the interaction Lagrangian of \autoref{eq:lagrangian}. The outer crust is tabulated using the mass table of Duflo and Zuker \cite{Duflo:1995} up to neutron drip $\sim 10^{-4}$ fm$^{-3}$. The outer crust and core equations of state are then interpolated using a cubic polynomial ensuring both the density and sound speed are continuous at each interface as in \cite{Piekarewicz:2019ahf}. We should note that the prescription for the inner crust may account for as much as 0.5km difference depending on the formalism used. However, the tidal deformability is insensitive to the crust EOS \cite{Piekarewicz:2019ahf}. The inner crust-core transition density is determined using the RPA-dynamical method \cite{Baym:1974vzp,Carriere:2002bx} with a slight modification to accompany the presence of the delta meson \cite{reed:thesis}. 

\begin{figure}[tb]
    \includegraphics[width=\columnwidth]{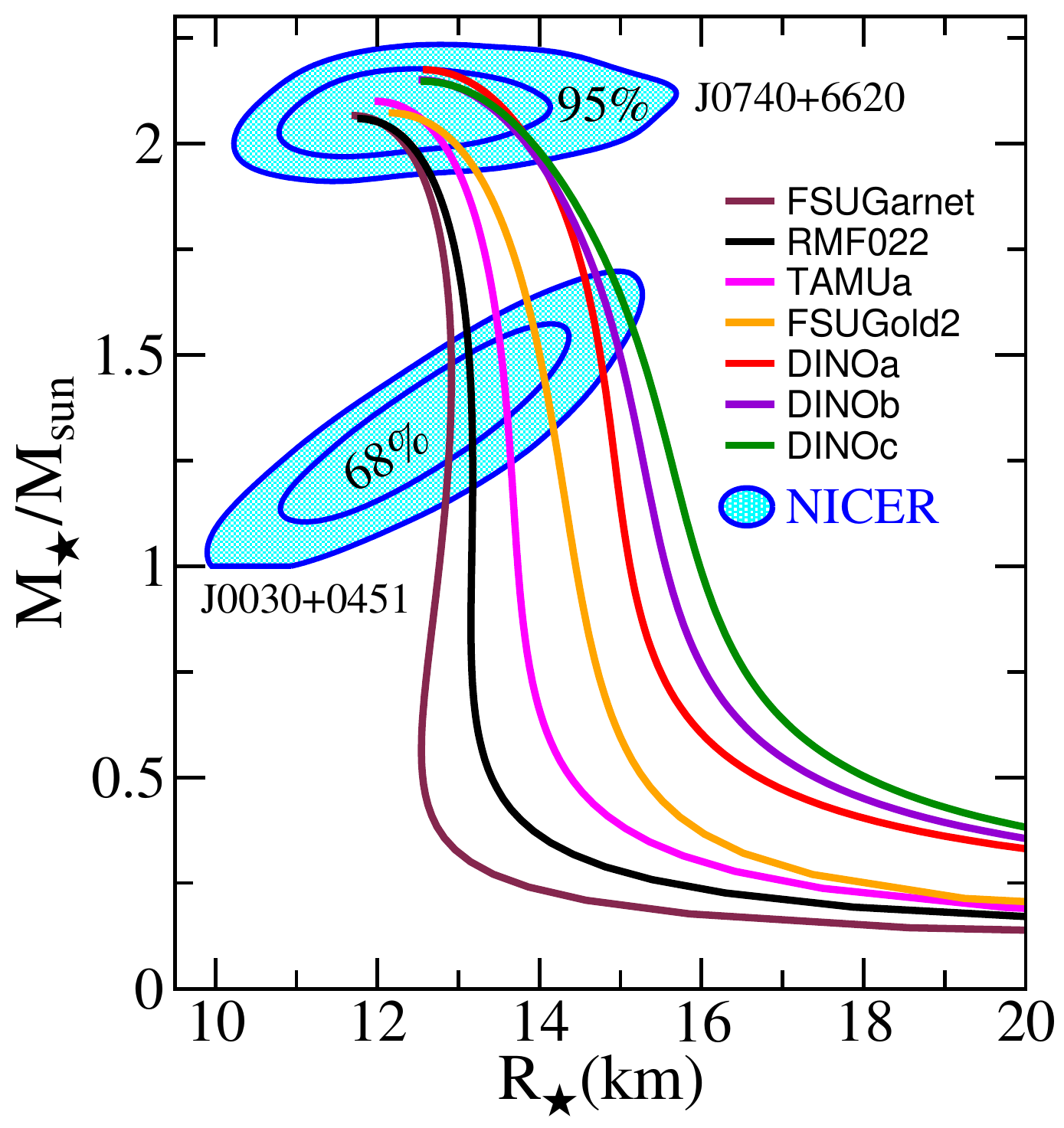}
    \caption{Mass-radius relation for neutron stars. Along with our three DINO models, predictions are also shown
    for a few covariant energy density functionals. The two sets of contours display the NICER mass-radius constraints of 
    two pulsars, PSR J0030+0451\,\cite{Miller:2019cac,Riley:2019yda} and PSR J0740+6620\,\cite{Miller:2021qha,Riley:2021pdl} at 68\% and 95\% confidence.}
    \label{fig:NS_mr}
\end{figure}

Although the DINO models are consistent with the properties of finite nuclei (see Table\,\ref{tab:BE_RC}) their large values for $\Ksym$ induce a dramatic stiffening of the EOS at the high densities of relevance to neutron stars. Such a stiffening generates neutron star radii that are too large when compared against LIGO-Virgo and NICER data. This behavior is illustrated in Fig.\ref{fig:NS_mr}, with explicit values in \autoref{tab:ns}, which displays the so called ``Holy Grail'' of neutron star physics: the mass-radius relation. Together with the predictions from the three DINO models and a few other covariant EDFs, the figure shows mass-radius determinations by the NICER mission for the two pulsars PSR J0030+045\,\cite{Miller:2019cac,Riley:2019yda} and PSR J0740+6620\,\cite{Miller:2021qha,Riley:2021pdl}. 

With radii of about 15\,km, the predictions from all three DINO models are inconsistent with the NICER data for the low mass star. With such large radii, the models also predict very large tidal deformabilities that are highly disfavored by GW170817\,\cite{Abbott:PRL2017,Malik:2018zcf,Most:2018hfd,Radice:2018ozg,Tews:2018chv}. In particular, we find that DINO models predict larger neutron star radii and lower densities for the direct Urca threshold. To agree with radii observation, one expects that a phase transition or other strong density dependence develops at intermediate densities. Moreover, a very small density for the direct Urca threshold suggest that all neutron stars with normal matter would undergo fast cooling which does not agree with to the cooling observations\,\cite{Beznogov:2015ewa}. If these predictions stand true, it raises the intriguing possibility of a neutron star core permeated with superfluid or superconductive matter throughout. Or in conjunction with the radii observation, this may even imply the prospect of an earlier onset of a phase transition to exotic states. Hence, we must conclude that while the DINO models provide a plausible solution to the PREX-2--CREX dilemma, they fail to reproduce astrophysical observations---unless a phase transition or other strong density dependence develops at intermediate densities. The emergence of a phase transition---normally accompanied by a softening of the EOS---has been shown to reduce the radius and deformability of neutron stars\,\cite{Chatziioannou:2021tdi,Han:2018mtj,Xie:2020rwg}.

\begin{table}[htb]
    \centering
    \begin{tabular}{|c|c|c|c|c|c|c|}\hline
    Model & $M_{\rm max}$ & $R_{1.4}$ & $\Lambda_{1.4}$ & $\rho_{\rm t}$ & $\rho_{\rm Urca}$ & $M_{\rm Urca}$\\
    & (M$_\odot$) & (km) & & (fm$^{-3}$) & (fm$^{-3}$) & (M$_\odot$)\\\hline
DINOa & 2.17 & 14.82 & 1050.6 & 0.0914 & 0.1580 & 0.418\\
DINOb & 2.15 & 15.11 & 1128.2 & 0.0846 & 0.1438 & 0.427\\
DINOc & 2.14 & 15.41 & 1240.4 & 0.0789 & 0.1373 & 0.458\\\hline
    \end{tabular}
    \caption{Prediction of various neutron star observables for our models. We calculate the core-crust transition density as described in \cite{Carriere:2002bx} with the modification to account for the addition of $\delta$-mesons.}
    \label{tab:ns}
\end{table}

\section{Conclusions}
\label{sec:conclusions}

In conclusion, the recently measured weak skin form factor in ${}^{48}$Ca by the CREX collaboration paints a 
peculiar picture for the behavior of neutron rich matter. When combined with the PREX-2 result that favors a
large value for $L$, the newly-revealed correlation between $F_{\rm Wskin}^{48}$ and the combination of 
bulk symmetry energy parameters $K_{\rm sym}\!-6L$, motivates the calibration of a new set of covariant 
EDFs with large and positive values for $K_{\rm sym}$. This result is in striking contrast to most non-relativistic 
EDFs and ab initio models that instead predict negative values for $K_{\rm sym}$. Nevertheless, the newly created 
DINO models with such unconventional values for $K_{\rm sym}$ agree better with PREX and CREX than many EDFs existent 
in the literature. However, this accomplishment comes at the price of stiffening the equation of state at the high 
densities of relevance to neutron stars, resulting in a failure to accommodate the recent NICER constraints. Nonetheless, these models represent a first step in reconciling PREX and CREX for covariant EDFs.

Moving forward, the Mainz Radius Experiment (MREX) at the future Mainz Energy-recovery Superconducting 
Accelerator (MESA)\,\cite{Becker:2018ggl} promises to increase the precision of $R_{\rm skin}^{208}$ by a factor 
of two. MREX will then confirm whether the slope of the symmetry energy is indeed stiff, or if the PREX-2 
measurement represents a large statistical fluctuation. Further, that most theoretical models disfavor large
and positive values for $K_{\rm sym}$ may suggest that the strong correlation uncovered here may be model 
dependent. Should this be the case, then ab initio models could inform how to improve the isovector sector of 
energy density functionals\,\cite{CREX:2013}. Finally, although the stiff symmetry energy suggested by the 
DINO models at high densities are ruled out by observation, these models could still be fruitful in describing
the properties of atomic nuclei---especially if a phase transition emerges at intermediate densities. Regardless, 
the PREX-2--CREX tension continues to reenergize the theoretical, experimental, and observational communities 
in our common quest to understand the behavior of dense, neutron rich matter.

\medskip
\begin{acknowledgments}\vspace{-10pt}
 We would like to thank Anna Watts for providing us with the NICER contours.
 This material is based upon work supported by the U.S. Department of Energy Office of Science, National Nuclear Security Administration of U.S. Department of Energy (Contract No. 89233218CNA000001), the Laboratory Directed Research and Development program of Los Alamos National Laboratory under project number 20230785PRD1, Office of Nuclear Physics under Awards DE-FG02-87ER40365 (Indiana University), Number DE-FG02-92ER40750 (Florida State University), and Number DE-SC0008808 (NUCLEI SciDAC Collaboration).
\end{acknowledgments} 

\bibliography{ReferencesJP}

\end{document}